\documentclass[10pt,oneside,twocolumn,aps,pra,preprintnumbers,bibnotes,superscriptaddress,amsmath,amssymb]{revtex4-2}

\usepackage[utf8]{inputenc}
\usepackage{graphicx} 
\usepackage{subfig}
\usepackage{caption}
\usepackage{epstopdf}
\usepackage{amsmath}
\usepackage{amssymb}
\usepackage{amsfonts}
\usepackage{amsthm}
\usepackage{mathtools}
\usepackage{bm}
\usepackage{braket}
\usepackage{physics}
\usepackage{cancel}
\usepackage[usenames,dvipsnames]{xcolor}
\usepackage{hyperref}
\usepackage{lipsum}
\usepackage{comment}
\usepackage{tikz}
\usepackage{tikzscale}
\usetikzlibrary{quantikz2,calc,arrows,chains,matrix,positioning,scopes}
\usepackage{letltxmacro}
\LetLtxMacro{\originaleqref}{\eqref}
\renewcommand{\eqref}{Eq.~\originaleqref}

\begin{document}

\title{Defect-induced stabilization of Wigner negativity and parity-protected phase-locking}

\author{Beatrice Donelli}
\affiliation{Istituto Nazionale di Ottica del Consiglio Nazionale delle Ricerche (CNR-INO), Largo Enrico Fermi 6, I-50125 Firenze, Italy.}
\affiliation{European Laboratory for Non-linear Spectroscopy, Universit\'a di Firenze, I-50019 Sesto Fiorentino, Italy.}

\author{Stefano Gherardini}
\affiliation{Istituto Nazionale di Ottica del Consiglio Nazionale delle Ricerche (CNR-INO), Largo Enrico Fermi 6, I-50125 Firenze, Italy.}
\affiliation{European Laboratory for Non-linear Spectroscopy, Universit\'a di Firenze, I-50019 Sesto Fiorentino, Italy.}

\begin{abstract}
We show that a parity-preserving delta-shaped potential, mimicking the presence of a localized defect, reduces the continuous $U(1)$-symmetry of a quantum van der Pol oscillator to the discrete $\mathbb{Z}_2$-symmetry. This mechanism guarantees protected even-parity stationary quantum coherences, resulting in the generation of non-classical steady-states characterized by Wigner function negativity. Thanks to a non-trivial interplay between the defect and non-linear dissipative channels, this kind of negativity entails parity-protected quantum phase locking, meaning that the oscillator's energy levels simultaneously lock into two opposite phases. Finally, we present a realistic experimental proposal involving the motional state of ultracold alkaline-earth-like atoms in optical tweezers, where the localized defect is engineered via state-selective contact interactions, while the non-linear dissipative channels are driven by Raman sidebands.
\end{abstract}

\maketitle

\section{Introduction}

The preparation of non-classical stationary quantum states is a central goal of quantum science and technologies~\cite{PoyatosCiracZoller1996, LeghtasScience2015, OfekNature2016, HarringtonNatureReview2022}. In continuous-variable quantum systems, non-classicality is witnessed by negative regions in the Wigner function (WF) of the system's state, which relate to quantum interference patterns~\cite{Wigner1932, Hudson1974, Kenfack2004, WeedbrookReviewGaussian}. However, the uncontrolled interaction with the external environment generally suppresses quantum coherences of the system's time-evolved state, which are responsible for WF negativity~\cite{BreuerPetruccione, Carmichael1993, GardinerZoller2004, Diehl2008}.
Therefore, achieving non-classical stationary states with negative WF is a challenging task, which so far has relied on finding the equilibrated competition between a coherent driving field and non-linear dissipative channels~\cite{Sieberer2013, NavarreteBenlloch2017}.

The quantum van der Pol oscillator (QvdPo) is a paradigmatic driven-dissipative system~\cite{vanderPol1926,Lee2013,Walter2014,Lorch2016,BenArosh2021}, which exhibits stable quantum limit cycles and lies at the heart of quantum synchronization~\cite{strogatzBOOK1994nonlinear,PikovskyBOOK2001,solanki2026reviewQS,Schmolke2026reviewQS,LahiriGupta2026}. The incoherent single-photon gain and nonlinear two-photon loss of the QvdPo generate a stationary state with continuous phase-rotation $U(1)$-symmetry. In the absence of phase-selective perturbations, this state is diagonal in the Fock basis and has a rotationally invariant, positive WF~\cite{Lee2013,Dodonov1997,Lorch2016}.
To break this continuous symmetry and stabilize quantum states with a locked phase, several strategies have been developed~\cite{Walter2014, Sonar2018, Lorch2016, Liu2026, Mok2020}. Among these, coherent harmonic driving provides an external phase reference inducing phase-locking~\cite{Walter2014}. To counteract quantum noise and improve phase synchronization, Sonar \textit{et al.}~\cite{Sonar2018} proposed replacing standard harmonic drives with parametric squeezing that allows for a pitchfork bifurcation into a symmetric two-lobe state. Instead, L\"{o}rch \textit{et al.}~\cite{Lorch2016} demonstrated that quantized multiphoton resonances, entailing steady-state WF negativity, can be generated through Kerr anharmonicities. More recently, field-parity engineering has been introduced to stabilize Wigner negativity by trapping excitations in odd-parity Fock states through multiphoton driving and dissipation~\cite{Liu2026}.

Nevertheless, a fundamental physical limitation arises in the deep quantum limit given by a large nonlinear two-photon dissipation strength: Mok \textit{et al.}~\cite{Mok2020} showed that squeezing decouples from the oscillator as two-photon loss restricts excitations to be near the vacuum state. This is because parametric squeezing is a two-photon process that requires population in the second-excited state. Consequently, squeezing-enhanced phase-locking is no longer active when two-photon loss dominates. Furthermore, the introduction of a coherent driving field is energetically unfavorable, apart from the fact that it requires precise time-dependent control function involving even highly non-linear modulations. Therefore, it would be desirable to have an autonomous, time-independent mechanism enabling the stabilization of the WF negativity, a quantum resource, whilst eliminating the need for coherent driving.

In this article, we demonstrate the generation and steady-state stabilization of quantum coherences in a QvdPo subjected to a time-independent, parity-preserving delta-shaped potential, which physically represents a localized defect. Unlike time-dependent parametric drives that are costly and fail near the vacuum, a spatially-localized defect directly reduces the continuous $U(1)$-symmetry to a discrete $\mathbb{Z}_2$ parity-symmetry. The defect couples even-parity Fock states and stabilizes even-parity coherences, specifically the lowest-order coherence $\rho_{02}$. Instead, elements of the oscillator's density operator involving odd-parity states vanish (thus, $\langle\hat a\rangle=0$) because their wave functions are zero at the origin of the phase space.

\begin{figure*}[t]
    \centering
    \includegraphics[width=0.33\linewidth]{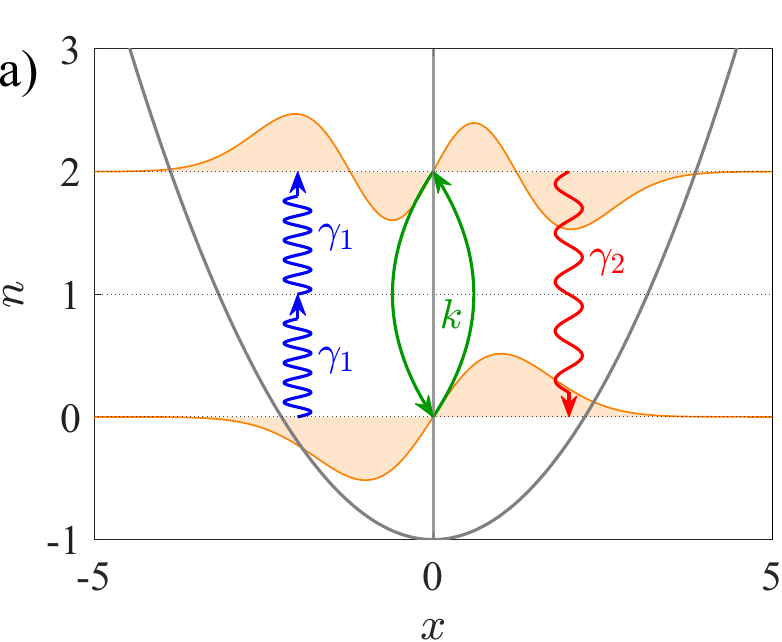}
    \hspace{0.005\linewidth}
    \includegraphics[width=0.32\linewidth]{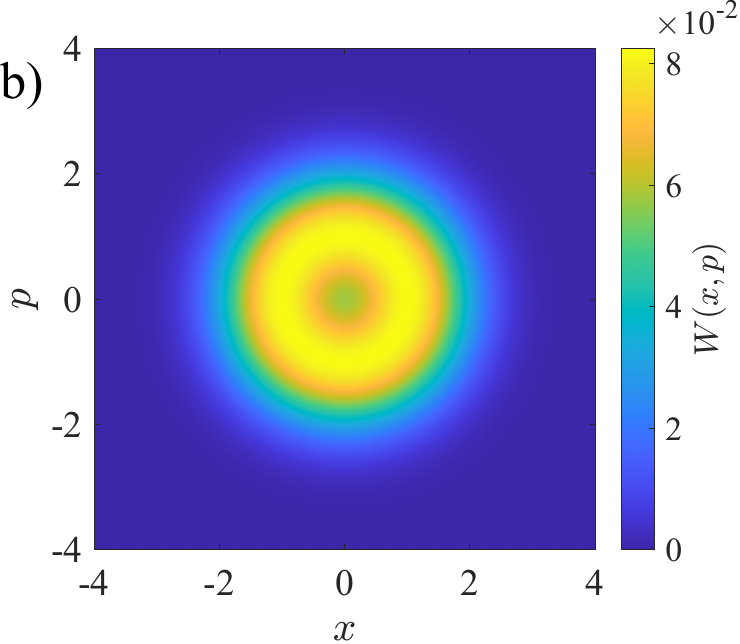}
    \includegraphics[width=0.32\linewidth]{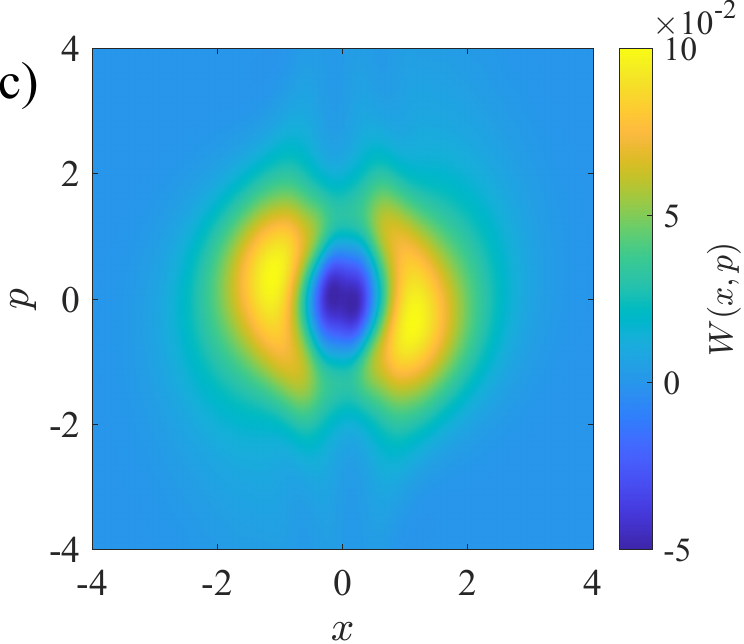}
    \caption{
    (a) Schematic representation of a QvdPo interacting with a delta-shaped potential. Here, we illustrate the first two even-parity eigenstates, corresponding to $n=0$ and $n=2$ (orange shaded areas). The eigenstate with $n=1$ is not shown, as the parity-protected symmetry separates the even- and odd-parity sectors. The figure also depicts the incoherent single-photon pumping with rate $\gamma_1$ (blue oscillating arrows), the incoherent two-photon dissipation with rate $\gamma_2$ (red oscillating arrows) and the coherent coupling induced by the potential with magnitude $k$ (green arrows). 
    (b) Stationary WF in the absence of the potential ($k=0$) with $\omega=1$, $\gamma_1 = 0.1\omega$ and $\gamma_1/\gamma_2 = 0.6$, showing the limit cycle with $U(1)$-symmetry. 
    (c) Stationary WF with the same parameters as panel (b), but in the presence of the potential with $k=10$, outlining the breaking of the $U(1)$-symmetry and the emergence of Wigner negativity originated by parity-protected coherences.
    }
    \label{fig:QvdP_scheme}
\end{figure*}

In the deep quantum limit, the lowest-order coherence $\rho_{02}$ dominates the phase-space interference pattern. It produces two phase-localized lobes and, above a strength threshold, guarantees negative regions in the WF at the steady-state~\cite{Kenfack2004,solinas2026operational}. Notably, the same lowest-order coherence $\rho_{02}$ also produces parity-protected quantum phase-locking, measured by the parity-invariant moment $\langle\hat a^2\rangle$~\cite{Mok2020, solanki2026reviewQS}.  
Consequently, a single defect modeled by a delta-shaped potential can be simultaneously responsible for WF negativity and parity-protected second-harmonic phase-locking, proving that quantum non-classicality and synchronization are dual manifestations of the same symmetry-breaking mechanism.

This defect-based mechanism could be realized using ultracold atomic mixtures, where a pinned impurity acts as an effective zero-range scatterer for a trapped atom, while Raman-transition sidebands and optical repumping engineer the single-photon gain and two-photon losses.

\section{Model and symmetries}
\label{sec:Modelandsymmetries}

We consider a driven-dissipative quantum oscillator, undergoing QvdPo dynamics~\cite{Lee2013,Walter2014,Lorch2016,BenArosh2021,Li2025_Exact}, which is additionally perturbed by a delta-shaped potential with magnitude $k$ (see Fig.~\ref{fig:QvdP_scheme}a). The time-evolution of this oscillator is described by the density operator $\hat\rho$, which is governed by the Gorini-Kossakowski-Sudarshan-Lindblad master equation ($\hbar = 1$)
\begin{equation}
    \frac{d\hat\rho_t}{dt} = \mathcal{L}_k[\hat\rho_t] :=  -i [ \hat{H}_k, \hat\rho_t ] + \gamma_1 \mathcal{D}[\hat{a}^\dagger]\hat\rho_t + \gamma_2 \mathcal{D}[\hat{a}^2]\hat\rho_t,
    \label{eq:master_eq}
\end{equation}
where $\mathcal{L}_k[\hat\rho]$ is the Liouvillian associated to the system's dynamics, $\hat{a}$ and $\hat{a}^{\dagger}$ are the annihilation and creation operators respectively, and we have made explicit the dependence of $\hat{H}$ on $k$. Here, $\gamma_1$ is the single-photon gain rate, $\gamma_2$ is the nonlinear two-photon dissipation rate and $\mathcal{D}[\hat{L}]\hat\rho := \hat{L}\hat\rho \hat{L}^\dagger - \frac{1}{2} \{\hat{L}^\dagger \hat{L}, \hat\rho \}$ is the Lindblad dissipator. The Hamiltonian of the QvdPo coupled to a defect localized around the position operator $\hat{x} =(\hat{a}+\hat{a}^\dagger)/\sqrt{2\omega}$ is
\begin{equation}
    \hat{H}_k = \omega \hat{a}^\dagger \hat{a} + k \delta(\hat{x}),
\end{equation}
where $\omega$ is the oscillator frequency.

In the absence of the potential ($k=0$), the Liouvillian $\mathcal{L}_0[\hat\rho]$ has a continuous $U(1)$-symmetry generated by the phase rotation $\hat{U}_\theta := e^{i\theta \, \hat{a}^\dagger \hat{a}}$. This is because both the Hamiltonian term $\omega \hat{a}^\dagger \hat{a}$ and the Lindblad dissipators are invariant under the transformation $\hat{a} \rightarrow e^{-i\theta}\hat{a}$. The steady state $\hat\rho_{\infty}^{\rm (k=0)}$ of the QvdPo therefore preserves this rotational symmetry of the Liouvillian $\mathcal{L}_0[\hat\rho]$ and is diagonal in the Fock basis: $\hat\rho_{\infty}^{\rm (k=0)} = \sum_{n} p_n |n\rangle\!\langle n|$. The corresponding WF is rotationally invariant and strictly positive, consistent with the classical limit-cycle structure of the QvdPo~\cite{Lee2013}.

The introduction of the delta-shaped potential ($k>0$) breaks the continuous $U(1)$-symmetry. However, since the potential is even under spatial inversion, i.e.~$\delta(\hat{x}) = \delta(-\hat{x})$, the system retains a discrete $\mathbb{Z}_2$ parity-symmetry generated by the parity operator $\hat{\Pi} \equiv \hat{U}_\pi = e^{i\pi \hat{a}^\dagger \hat{a}}$. 
The Liouvillian $\mathcal{L}_k[\hat\rho]$ is parity symmetric, namely $\mathcal{L}_k[\hat{\Pi}\hat{\rho}\hat{\Pi}^\dagger]=\hat{\Pi}\mathcal{L}_k[\hat{\rho}]\hat{\Pi}^\dagger$. In fact, the parity operator transforms the ladder and position operators as $\hat{\Pi}\hat{a}\hat{\Pi}^\dagger=-\hat{a}$ and $\hat{\Pi}\hat{x}\hat{\Pi}^\dagger=-\hat{x}$, respectively, leaving the Hamiltonian unchanged. Similarly, the Lindblad dissipators $\mathcal{D}[\hat{a}^\dagger]$ and $\mathcal{D}[\hat{a}^2]$ are invariant under parity. Consequently, the steady state of the QvdPo commutes with the parity operator: $[\hat{\Pi}, \hat\rho_{\infty}] = 0$.

Because $\hat{a}$ is odd under parity transformations, its expectation value vanishes:  
\begin{equation}
    \langle \hat{a} \rangle = \text{Tr}[\hat{a} \hat{\rho}_{\infty}] = -\text{Tr}[\hat{\Pi} \hat{a} \hat{\Pi}^\dagger \hat{\rho}_{\infty}] = -\langle \hat{a} \rangle = 0,
\end{equation}
thus fixing the center of the WF at the origin of the phase-space. While parity symmetry suppresses odd moments, it allows coherences between states of equal parity: $\bra{n} \hat{\Pi} \hat{\rho}_{\infty} \hat{\Pi}^\dagger \ket{m} = (-1)^{n+m} \rho_{\infty}^{(n,m)}$, where $\rho_{\infty}^{(n,m)} := \bra{n}\rho_{\infty}\ket{m}$. As a result, only elements with even $n+m$ survive, giving the density operator a checkerboard structure in the Fock basis. The localized defect creates even-parity coherences while forbidding odd ones, suggesting a fundamental parity-symmetry mechanism, which we are going to explain below. These protected coherences are responsible for the emergence of the non-classical features observed in Fig.~\ref{fig:QvdP_scheme}c, where the system is forced into a superposition of phase-localized states. The latter is associated to a negative WF, as it can observed from panels b)-c) of Fig.~\ref{fig:QvdP_scheme}. In the absence of the potential ($k=0$), the WF exhibits a characteristic ring shape, representing the rotationally invariant limit-cycle of the unperturbed QvdPo. When the defect is coupled to the QvdPo ($k > 0$), the $U(1)$-symmetry of the oscillator is broken down to $\mathbb{Z}_2$, concentrating the WF into two distinct lobes. The interferences between these two lobes lead to regions of negativity in the WF (dark blue areas in panel c), which represents a quantum feature of the perturbed system absent in the $k=0$ case. 

\section{Exact steady-state solution}

The potential $\delta(\hat{x})$ introduces a {\it nonlocal} coupling between levels of the Fock space. A key feature arises from the fact that the perturbed QvdPo's eigenfunctions have definite parity, which implies that $\delta(\hat{x})$ only couples even Fock states. 
To make this structure explicit, we introduce the oscillator's state at the origin of the position-momentum phase space, $|\psi(0)\rangle = \sum_{n \ \mathrm{even}} \psi_n(0)\,|n\rangle$, where $\psi_n(0) := \langle n|x=0\rangle$. In this representation, the potential takes the form of a projection onto the origin $x=0$:
\begin{equation}
    \delta(\hat{x}) = |\psi(0)\rangle \langle\psi(0)|.
\end{equation}

Before addressing the solution of the perturbed QvdPo's dynamics, we recall the exact steady-state solution in the unperturbed case $(k=0)$. Thanks to the system's symmetries, the steady-state is diagonal in the Fock basis: $\hat\rho_{\infty}^{\rm (k=0)} = \sum_n p_n |n\rangle\langle n|$. The populations $p_n$ satisfy a recurrence relation derived from $\mathcal{L}_0[\hat\rho_{\infty}^{\rm (k=0)}]=0$ and the corresponding solution reads as~\cite{Dodonov1997, Lorch2016, Li2025_Exact}:
\begin{equation}\label{eq:pn_unpert}
    p_n = \frac{\frac{\gamma_1^n}{\gamma_2^n}}{(\gamma_1/\gamma_2)_n} \frac{\Phi\left(1+n; \frac{\gamma_1}{\gamma_2}+n; \frac{\gamma_1}{\gamma_2}\right)}{\Phi\left(1; \frac{\gamma_1}{\gamma_2}; \frac{2\gamma_1}{\gamma_2}\right)},
\end{equation}
where $(\cdot)_n$ is the Pochhammer symbol and $\Phi(a;b;z)$ is the Kummer confluent hypergeometric function (see Appendix \ref{app:unperturbed_exact_solution} for details).

For $k>0$, the steady-state equation $\mathcal{L}_k[\hat\rho_{\infty}] = 0$ is equivalent to
\begin{equation}\label{eq:steady-state-eq}
    \mathcal{L}_0[\hat\rho_{\infty}] = ik \left[ \delta(\hat{x}), \hat\rho_{\infty} \right].
\end{equation}
To solve \eqref{eq:steady-state-eq}, we introduce the Green's super-operator $\mathcal{G} := \mathcal{L}_0^{+}$, defined as the Moore-Penrose pseudoinverse of the unperturbed Liouvillian. This generalized inverse is required because $\mathcal{L}_0$ is singular, having a zero eigenvalue associated with the unperturbed steady state $\hat{\rho}_\infty^{\rm (k=0)}$. Applying $\mathcal{G}$ to \eqref{eq:steady-state-eq} yields a formal solution analogous to the Lippmann-Schwinger equation of scattering theory~\cite{Lippmann1950,Weinberg1995}:
\begin{equation}\label{eq:full_steady-state_problem}
    \hat\rho_{\infty} = \hat{\rho}_\infty^{\rm (k=0)} + ik\, \mathcal{G}\Big[ [\delta(\hat{x}), \hat\rho_{\infty}] \Big],
\end{equation}
where we have used the identity $\mathcal{G}\mathcal{L}_0[\hat{\rho}_\infty]=\hat{\rho}_\infty - \hat{\rho}_\infty^{(k=0)}$. The super-operator $\mathcal{G}$ is independent on the steady-state $\hat{\rho}_\infty$ of the perturbed system, but depends on the parameters $(\omega,\gamma_1,\gamma_2)$. By defining the auxiliary vector $|R\rangle := \hat\rho_{\infty}|\psi(0)\rangle$, the steady-state problem of \eqref{eq:full_steady-state_problem} reduces to a system of equations for the coefficients of $|R\rangle$, which we expand in the even Fock basis as $|R\rangle = \sum_{m \,\rm even} r_m |m\rangle$. We thus obtain a system in the vector $\vec{r} := (r_1,\ldots,r_m,\ldots)^T$: 
\begin{equation}
    \vec{r} = \vec{r}_0 + ik B\, \vec{r}^{\,*} - ik A\, \vec{r},
    \label{eq:solution_with_r}
\end{equation}
where $\vec{r}_0 := (s_1,\ldots,s_m)^T$ is the vector of coefficients that compose $|R_0\rangle := \hat\rho_{\infty}^{\rm (k=0)}|\psi(0)\rangle = \sum_{m \,\rm even} s_m |m\rangle$, which are known and $(\cdot)^{*}$ denotes complex conjugation. Moreover, the matrices $A$ and $B$, representing the action of the Green's superoperator, are defined as
\begin{eqnarray}
    B_{jm} &:=& \langle j| \, \mathcal{G}\Big[ |\psi(0)\rangle\langle m| \Big]  |\psi(0)\rangle  \\
    A_{jm} &:=& \langle j| \, \mathcal{G}\Big[ |m\rangle\langle\psi(0)| \Big] |\psi(0)\rangle.
\end{eqnarray}
Such a formulation reduces the dimension of the numerical problem from $O(N^2)$, which is the size of the Liouvillian, to $O(N)$ that is the size of $\vec{r}$. Solving \eqref{eq:solution_with_r} allows us to explore the structure of the steady-state $\hat{\rho}_\infty$ across different regimes of the potential magnitude $k$ and dissipation rates $\gamma_{1}, \gamma_{2}$. In the weak-perturbation limit ($k \ll \omega, \gamma_1, \gamma_2$), $\hat{\rho}_\infty$ still exhibits the $U(1)$-symmetric ring in the phase space, while for large $k$ the defect acts as a hard barrier at the origin. This transition from a continuous limit cycle to a parity-protected bipartition can be well-visualized by looking at the WF of the perturbed steady state, which we are going to analyze below.

\section{Stationary Wigner function negativity}
\label{sec:WF_negativity}

In the unperturbed QvdPo ($k=0$), the steady state $\hat\rho_{\infty}^{\rm (k=0)}$ is diagonal in the Fock basis, resulting in a rotationally symmetric and strictly positive WF (see Fig.~\ref{fig:QvdP_scheme}b). The coupling to a localized defect, modeled by a delta-shaped potential, reduces the continuous $U(1)$-symmetry to the discrete $\mathbb{Z}_2$ parity-symmetry. The latter strictly forbids odd-parity elements of the system's density matrix (i.e., $\rho_{nm}=0$ for $n+m$ odd), thus the defect induces a global coupling between even-parity levels. Consequently, the steady state $\hat\rho_{\infty}$ develops off-diagonal coherences between states differing by an even number of excitations. These parity-protected coherences manifest as an anisotropic interference pattern in the position-momentum phase space.

Interestingly, the lowest-order quantum coherence $\rho_{02} = |\rho_{02}|e^{i\phi_{02}}$ plays the dominant role in the emergence of the WF negativity, a witness of non-classicality. This is particularly evident in the deep quantum limit, which is characterized by a small gain-to-loss ratio: $\gamma_1/\gamma_2 \ll 1$. In this deep quantum limit, the limit-cycle structure supporting $U(1)$-symmetry in the WF of the QvdPo is localized near the vacuum. This means that the potential perturbation involves only a few system's energy levels, in addition to the ground state. Moreover, if the coherent coupling induced by the potential is sufficiently weak compared to the two-photon dissipation (i.e., $\gamma_2$ dominates over both $\gamma_1$ and $k$), then the populations of the QvdPo's state remain close to their unperturbed values; see also Appendix~\ref{app:truncation_calculations}. As a result, because all higher levels are strongly suppressed as $p_{n\ge 3} \sim \mathcal{O}(\gamma_1^2/\gamma_2^2)$, the transition from the $U(1)$-symmetry to the discrete $\mathbb{Z}_2$ parity-symmetry is well-captured analytically by truncating $\hat\rho_{\infty}$ to the manifold of levels with $n\leq 2$.

In order to rigorously justify this truncation, in Appendix~\ref{app:Truncation} we compare the exact steady state $\hat\rho_{\infty}$ with the corresponding approximation within the $n\leq 2$ manifold, using the trace distance. As expected, the trace distance grows when high-excited Fock states are taken into account, thus for $\gamma_1/\gamma_2 \gtrsim 1$. We have also determined the relative weight of $\rho_{02}$ with respect to the $\ell_1$-norm of coherence as a function of $k$ and $\gamma_1/\gamma_2$, finding that its importance is greatest precisely when the trace distance is smallest.  Both these measures numerically confirm the quality of the approximation in the $n\leq 2$-truncated manifold to represent the steady state of the perturbed QvdPo in the deep quantum limit.

Within the truncated manifold, the stationary WF---denoted as $\widetilde{W}$---can be analytically computed. Defining the complex phase-space amplitude $\alpha := (x+ip)/\sqrt{2}$ and using the polar variables $u := 2|\alpha|^2$ and $\theta := \arg(\alpha)$, the expansion of $\widetilde{W}$ in terms of the Laguerre polynomials yields:
\begin{eqnarray}\label{eq:WignerExact}
    \widetilde{W}(u, \theta) = \frac{e^{-u}}{\pi} \Big[ & p_0 - p_1(1 - 2u) + p_2(1 - 4u + 2u^2)\nonumber \\
    &+ 2\sqrt{2}|\rho_{02}|u \cos(2\theta + \phi_{02}) \Big],    
\end{eqnarray}
where $p_0, p_1, p_2$ are the populations of the system's state in the first three Fock levels. \eqref{eq:WignerExact} reveals the physical origin of the WF negativity, signature of non-classicality. The terms in the first line of (\ref{eq:WignerExact}) depend solely on the populations and provide a purely radial, positive contribution to the WF. Instead, the leading-order coherence $\rho_{02}$ introduces angular interference fringes that can drive the WF below zero.

In phase space, the maximum destructive quantum interference occurs along the line identified by the condition $\cos(2\theta + \phi_{02}) = -1$. Evaluating the WF along these `valleys', we determine that the stationary WF reduces to a one-dimensional Gaussian-damped parabola: $\widetilde{W}_{\rm valley}(u) = \frac{e^{-u}}{\pi} (A + Bu + Cu^2)$ with coefficients
\begin{equation}
    A := p_0 - p_1 + p_2, \quad B := 2p_1 - 4p_2 - 2\sqrt{2}|\rho_{02}|, \quad C := 2p_2. 
\end{equation}
Since $C>0$ for $p_2\neq0$, the parabola has positive concavity. For the WF to dip below zero at a physically accessible radius $(u > 0)$, the minimum of the parabola must be negative. This geometric requirement is satisfied if the discriminant $\Delta = B^2 - 4AC$ is positive, which simplifies to the linear condition $B < -2\sqrt{AC}$ for the branch with $B<0$. This yields: 
\begin{equation}
    \sqrt{2} |\rho_{02}| > p_1 - 2p_2 + \sqrt{2p_2(p_0 - p_1 + p_2)} \,.
    \label{eq:NegativityCondition}
\end{equation} 
Equation~\ref{eq:NegativityCondition} reveals that Wigner negativity is achieved only when the parity-protected coherence $|\rho_{02}|$ exceeds a critical threshold set by the populations $\{p_n\}$. As $|\rho_{02}|$ is controlled by the coupling of magnitude $k$ with the defect, physically \eqref{eq:NegativityCondition} outlines the competition between the coherent coupling with the defect and the incoherent dissipative channels with rates $\gamma_1, \gamma_2$, within the manifold of levels $n\leq 2$. Dissipative channels bring the perturbed system towards a diagonal state, thus ``blurring'' the interference patterns generated by the potential, while the negativity of the WF emerges only when the coherent coupling is strong enough to prevent these patterns from being erased at the steady-state.

In order to relate the geometric condition of \eqref{eq:NegativityCondition} to the physical parameters $(\gamma_1, \gamma_2, k,\omega)$ of the model, we solve the perturbed steady-state of the QvdPo within the truncated manifold, whose populations $\{p_n\}$ and quantum coherence $|\rho_{02}|$ are functions of $(\gamma_1, \gamma_2, k, \omega)$ (see Appendix~\ref{app:truncation_calculations}). In this way, we can determine a closed-form analytical boundary for the critical potential magnitude, $k_{\rm crit}$, which is obtained from considering \eqref{eq:NegativityCondition} as an equality: 
\begin{equation}
    k_{\rm crit} \approx \sqrt{\frac{\pi}{\omega}} \sqrt{4\omega^2 + \Gamma_{02}^2} \left[ \frac{1 - 2\frac{\gamma_1}{\gamma_2} + \sqrt{2\frac{\gamma_1}{\gamma_2}(1+\frac{\gamma_1}{\gamma_2})}}{2-\frac{\gamma_1}{\gamma_2}} \right], 
    \label{eq:k_crit_analytical}
\end{equation}
where $\Gamma_{02} := 2\gamma_1+\gamma_2$. In the extreme deep quantum limit ($\gamma_1/\gamma_2 \to 0$), \eqref{eq:k_crit_analytical} simplifies to $k_{\rm crit} \approx \frac{1}{2}\sqrt{\frac{\pi}{\omega}(4\omega^2 + \gamma_2^2)}$. This analytical limit defines a phase boundary in the bi-dimensional parameter space defined by $(k/\omega, \gamma_1/\gamma_2)$, separating the positive, classical limit cycle from the non-classical regime characterized by Wigner negativity at the origin where the defect is located.

\begin{figure}[t]
    \centering
    \includegraphics[width=\columnwidth]{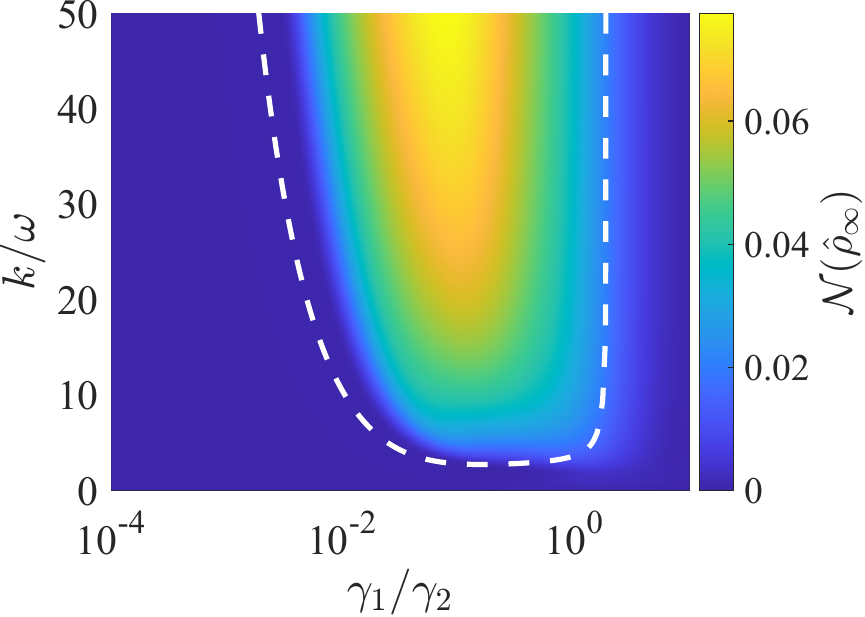}
    \caption{
    Phase diagram of the steady-state Wigner negativity $\mathcal{N}(\hat{\rho}_\infty)$ as a function of the normalized potential magnitude $k/\omega$ and the ratio $\gamma_1/\gamma_2$, with $\gamma_1 = 0.1\omega$. The white dashed line represents the analytical threshold $k_{\rm crit}$ [\eqref{eq:k_crit_analytical}] derived from the $n=2$ Fock-space truncation. Excellent agreement of this approximation is achieved in the deep quantum limit.
    }
    \label{fig:Negativity}
\end{figure}

The Wigner negativity $\mathcal{N}$ corresponds to the total volume of the negative regions of the WF~\cite{Kenfack2004, solinas2026operational, Rohith2023}. At the steady-state, we find:
\begin{equation}
    \mathcal{N}(\hat{\rho}_\infty) = \int \frac{ |\widetilde{W}(x,p)| - \widetilde{W}(x,p)}{2} dx dp \,.
\end{equation}
In Fig.~\ref{fig:Negativity}, we present the phase diagram of the exact steady-state Wigner negativity $\mathcal{N}(\hat{\rho}_\infty)$ (computed numerically without truncation) as a function of both $k/\omega$ and the gain-to-loss ratio $\gamma_1/\gamma_2$. Superimposed on the colormap is the analytical threshold $k_{crit}$ of \eqref{eq:k_crit_analytical} (white dashed line) derived from the ($n\leq 2$)-manifold approximation, which well confines the Wigner negativity. Interestingly, while the truncation at $n\leq 2$ is rigorously justified only in the deep quantum limit ($\gamma_1/\gamma_2 \ll 1$), the analytical threshold $k_{\rm crit}$ [\eqref{eq:k_crit_analytical}] successfully tracks the onset of Wigner negativity across a remarkably broad parameter regime. In the phase space, as the ratio $\gamma_1/\gamma_2$ increases, the limit cycle expands and the system's populations $\{p_n\}$ spread across high-energy Fock levels. As a consequence, a stronger potential magnitude $k$ is required to generate interference patterns of such a spread population background. The analytical formula for the threshold $k_{\rm crit}$ correctly captures this functional scaling up to $\gamma_1/\gamma_2 \approx 2$. At exactly $\gamma_1/\gamma_2=2$, the denominator in \eqref{eq:k_crit_analytical} causes an unphysical divergence, highlighting the explicit breakdown of the $n\leq 2$ approximation, beyond which the coupling with others higher-order manifolds can no longer be neglected.

Finally, we observe that the Wigner negativity is most pronounced when $k$ is large enough to split the limit cycle into two phase-localized lobes, as depicted in Fig.~\ref{fig:QvdP_scheme}c. The presence of these lobes at the steady-state is linked to the $\mathbb{Z}_2$-symmetry. In fact, any parity-breaking perturbation---such as displacing the potential away from the origin $x=0$---reintroduces quantum coherences between odd-parity levels leading to $\langle \hat{a} \rangle \neq 0$, which spoils the parity-protected interference patterns.

\section{Parity-protected quantum phase locking}

In conventional systems exhibiting limit-cycles in their phase space, including the standard QvdPo~\cite{Lee2013, Walter2014, Lorch2016, BenArosh2021, Li2025_Exact}, phase-locking is detected by the expectation value $\langle \hat{a} \rangle$ or by the normalized order parameter $S_1 := \langle \hat{a} \rangle / \sqrt{\langle \hat{a}^\dagger \hat{a} \rangle}$. A nonzero value of $S_1$ signals spontaneous breaking of the continuous $U(1)$-symmetry and the emergence of an in-phase correlation of system’s energy levels at the steady-state~\cite{Lee2013,Walter2014,solanki2026reviewQS}. In the presence of a localized defect, $\langle \hat{a} \rangle = 0$ [see also Sec.~\ref{sec:Modelandsymmetries}] such that $S_1$ vanishes identically. Nevertheless, as the perturbed system preserves the $\mathbb{Z}_2$ parity-symmetry, the appropriate order parameter to be chosen must be invariant under the parity transformation $\hat{a}\rightarrow -\hat{a}$. Therefore, the simplest nontrivial order parameter, accounting for the phase-locking under parity-symmetry, is~\cite{Sonar2018,Mok2020}
\begin{equation}\label{eq:SyncMeasure}
    S_2 = \frac{\langle \hat{a}^2 \rangle}{\langle \hat{a}^\dagger \hat{a} \rangle} 
    = \frac{{\rm Tr}\left[ \hat{\rho}_{\infty} \hat{a}^2 \right]}{ {\rm Tr}\left[ \hat{\rho}_{\infty} \hat{a}^\dagger \hat{a}\right]},
\end{equation}
which is sensitive to the presence of quantum coherence between energy levels differing by two excitations. Within the $n \leq 2$-truncated manifold, using the expression of the steady-state $\hat{\rho}_{\infty}$ described in Appendix~\ref{app:truncation_calculations}, the second-order parameter $S_2$ simplifies to
\begin{equation}\label{eq:synchronization_2}
    S_2 = \frac{ \langle 2| \hat{\rho}_{\infty} \hat{a}^2 |2 \rangle }{ p_1 + 2 p_2 } = \frac{3\sqrt{2} \gamma_2 }{\Gamma_{02}} \rho_{02}^{*}.
\end{equation}
It is worth noting that $S_2$ is directly proportional to the parity-protected coherence $\rho_{02}$, which is also responsible for stationary WF negativity in the analyzed parameter regimes (see Sec.~\ref{sec:WF_negativity}).

\begin{figure*}[t]
    \centering
    \includegraphics[width=0.45\linewidth]{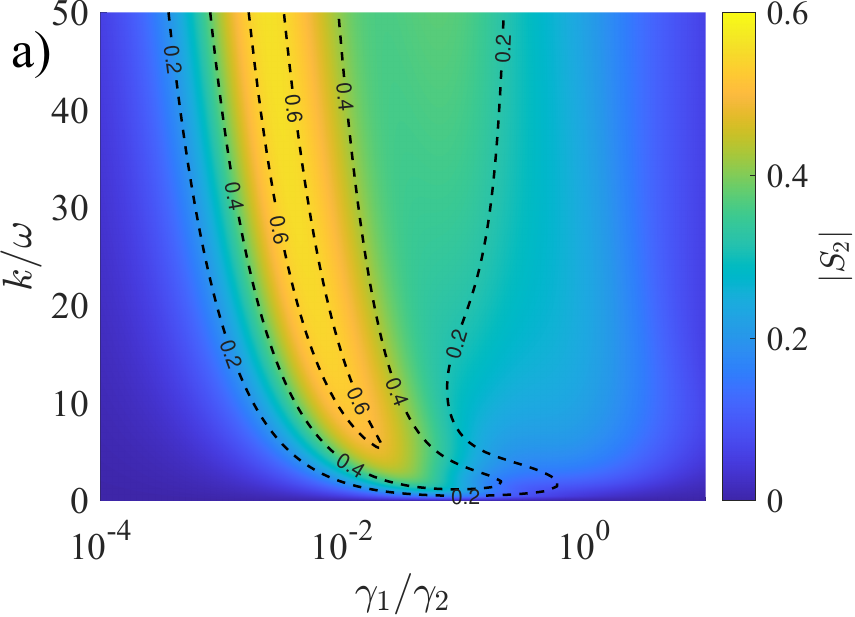}
    \includegraphics[width=0.45\linewidth]{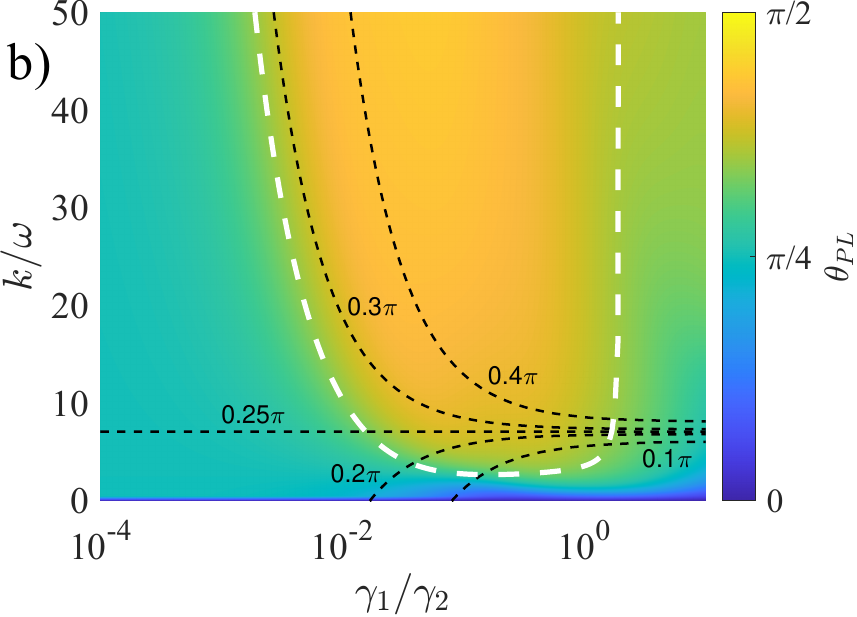}
    \caption{ 
    Steady-state parity-protected quantum phase locking as a function of the normalized potential magnitude $k/\omega$ and the gain-to-loss ratio $\gamma_1/\gamma_2$. 
    In both panels, the colormap displays the exact numerical steady-state solution, while the overlaid black dashed contours indicate the analytical values derived within the $n \leq 2$-truncated manifold. 
    (a) Modulus of the second-order phase-locking parameter $|S_2|$. 
    (b) Locking angle $\theta_{\text{PL}} = \frac{1}{2}\arg \langle\hat{a}^2\rangle$. 
    The region of maximum phase locking is well-limited by the same WF negativity boundary plotted in Fig. \ref{fig:Negativity} (white dashed line), where $\mathcal{N}(\hat{\rho}_\infty) > 0$. 
    }
    \label{fig:synchronization}
\end{figure*}

The modulus $|S_2|$ measures the magnitude of this parity-protected phase-locking mechanism, as shown in Fig.~\ref{fig:synchronization}(a). The in-phase correlation between system's levels in the Fock basis occurs in the deep quantum regime ($\gamma_1/\gamma_2 \ll 1$), and is well-captured by the black analytical contours derived from computing the modulus $|S_2|$ in the $n \leq 2$-truncated manifold. For the modulus $|S_2|$, the agreement between numerical simulations (which include all levels populated at the steady-state) and our analytical prediction using the truncated manifold is excellent.

The phase of $S_2$ defines the angle $\theta_{\rm PL}$ at which the WF has a locked-phase, corresponding to the orientation of the locked state generated by the coupling with the defect: 
\begin{equation}
    \theta_{\rm PL} = \frac{1}{2} \, \arg \langle \hat{a}^2 \rangle \quad \text{mod} \,\,\pi,
\end{equation}
where the factor $1/2$ reflects that the expectation value of $\hat{a}^2$ carries a phase $2\theta$ entailed by the parity-symmetry. Accordingly, $\theta_{\rm PL}$ is defined modulo $\pi$, which corresponds to two opposite locked phases. Interestingly, within the truncated manifold, the phase $\theta_{\rm PL}$ reduces to the phase of the quantum coherence $\rho_{02}^{*}$. Figure~\ref{fig:synchronization}(b) displays $\theta_{\rm PL}$ as a function of model parameters $k/\omega$ and $\gamma_1/\gamma_2$. This angle $\theta_{\rm PL}$ varies from $0$ to $\pi/2$ and it is maximized
in the same parameter region where the stationary WF of the perturbed system is negative (white dashed line). Our analytical predictions in the $n \leq 2$-truncated manifold capture the behavior of $\theta_{\rm PL}$ in the deep quantum regime ($\gamma_1/\gamma_2 \ll 1$), but clearly diverge from the numerical results as the gain-to-loss ratio increases, in correspondence with the breakdown of the validity of the ($n \leq 2$)-manifold approximation. Hence, within the ($n \leq 2$) manifold in the deep quantum limit, parity-protected phase-locking and Wigner negativity are intertwined manifestations of the same defect-induced continuous-symmetry breaking effect.

In order to complement the information provided by $|S_2|$, phase-locking can be also characterized via the angular phase distribution $P(\theta)$, obtained by integrating the WF over the radial direction~\cite{Walter2014,Li2025}:
\begin{equation}\label{eq:def_P_theta}
    P(\theta) = \int_0^\infty W(u,\theta) \, u \, du.
\end{equation}
Within the ($n\leq 2$)-manifold, substituting \eqref{eq:WignerExact} into the integral of \eqref{eq:def_P_theta} yields:
\begin{equation}
    P(\theta) = \frac{1}{2\pi}\left( 1+2\sqrt{2}|\rho_{02}|\cos{(2\theta + \phi_{02})} \right),
    \label{eq:P_phi}
\end{equation}
where we recall that $\phi_{02} = \arg(\rho_{02})$. The function $P(\theta)$ exhibits two symmetric maxima located at $\theta_1 = -\phi_{02}/2$ and $\theta_2 = \theta_1 + \pi$. These peaks match the locking angles $\theta_{\mathrm{PL}}$, as shown in Fig.~\ref{fig:synchronization}(b). 
The peak value of the distribution can be analytically computed within the $(n\leq 2)$-manifold approximation, and reads as
\begin{equation}
    \max_\theta P(\theta) = \frac{1 + 2\sqrt{2}|\rho_{02}|}{2\pi}.
\end{equation}

A natural counterpart to $S_2$ is the second-harmonic Mean Resultant Length (MRL) of a circular distribution $R_2 := |\langle \hat{\chi}^2 \rangle| = |\int_0^{2\pi} e^{i2\theta} P(\theta) d\theta| $~\cite{Li2025,Mok2020}, where $\hat{\chi}= \sum_{n=0}^\infty |n\rangle\!\langle n+1|$ is the phase operator formulated within the Pegg-Barnett formalism~\cite{PeggBarnett1989} (see Appendix~\ref{app:MRL} for details). 
This quantity offers a quantum analogue of the second-order classical Kuramoto order parameter~\cite{Kuramoto1984, StrogatzKuramoto2000, Gherardini2018} and has a clear physical interpretation: $R_2=0$ for an unsynchronized state, while $R_2=1$ for a phase-locking into two symmetric $\pi$-separated states. Within the $(n\leq 2)$-manifold, we have $\hat{\chi}^2 \approx |0\rangle\!\langle 2|$, yielding $R_2 \approx |\rho_{02}|$ and $|S_2| \approx \sqrt{2} R_2$. Therefore, $S_2$, $R_2$, and the peaks of $P(\theta)$ represent fully equivalent diagnostic quantifiers for parity-protected quantum phase-locking in the deep quantum limit.

Finally, phase-locking induced by the coupling with a localized defect can be also identified by the dynamical speed at which the system's state reaches the steady-state. Such a speed is governed by the Liouvillian spectral gap:
\begin{equation}
    \Delta_{\mathcal{L}} = -\max_{\lambda \neq 0} \mathrm{Re}(\lambda),    
\end{equation}
where $\lambda$ are the eigenvalues of the Liouvillian superoperator $\mathcal{L}_k$. In the unperturbed QvdPo ($k=0$), continuous $U(1)$-symmetry implies a phase diffusion along the limit cycle with a vanishing spectral gap. Instead, introducing a defect ($k > 0$) opens a finite gap, $\Delta_{\mathcal{L}} > 0$~\cite{Albert2014,Minganti2018}. As the timescale for achieving steady-state phase-locking is $\tau_{\mathrm{lock}} = 1/\Delta_{\mathcal{L}}$, increasing the interaction strength $k$ with the defect widens the Liouvillian gap and accelerates the relaxation towards the phase-locked state.

\section{Experimental proposal}

Defect-induced stabilization of Wigner negativity and parity-protected phase-locking in an autonomous setting could be observed with ultracold fermionic alkaline-earth-like atoms (e.g., $^{171}\text{Yb}$ or $^{87}\text{Sr}$) held in optical tweezers.

While trapped ions are a standard platform for engineered motional dissipations~\cite{PoyatosCiracZoller1996,Lee2013,HarringtonNatureReview2022,Li2025}, their tight spatial confinement ($x_{\rm zpf} \sim 10$ nm) precludes the realization of a true localized Dirac-delta defect below the optical diffraction limit. Instead, it is worth considering a single primary neutral atom confined in a shallow 1D optical dipole trap. This trap acts as the unperturbed harmonic oscillator with frequency $\omega \approx 2\pi \times 100\,\text{Hz}$, yielding a large zero-point spatial spread $x_{\rm zpf} \sim 1 \, \mu\text{m}$. To engineer the parity-preserving delta-shaped potential $k\delta(\hat{x})$ mimicking the defect, we follow the physical mechanism recently proposed in Refs.~\cite{DonelliSciPost2026,DonelliPRA2026}. Specifically, the defect is realized through an additional impurity-atom of the same alkaline-earth-like species, prepared in a different internal state. State-selective potentials and interatomic contact interactions~\cite{Jaksch2005,Recati2005} are used to generate the desired localized perturbation. 
The impurity-atom is pinned at the center of the primary atom's harmonic trap ($x=0$) using a tightly focused, state-selective optical tweezer ($\omega_{\text{imp}} \gg \omega$). 
In the low-energy limit, and within the zero-range approximation, the resulting interactions reduce to an effective single-particle dynamics subjected to a delta-like perturbation in a harmonic trap~\cite{Busch1998}, taking the form $V(\hat{x}) \approx g_{\rm 1D}\delta(\hat{x})$, where $g_{\rm 1D}$ denotes the perturbation strength~\cite{pethick_bose-einstein_2008, pitaevskii_bose-einstein_2016}. Tunable interactions between the primary atom and the impurity are achieved via inter-state scattering lengths between the $^1\text{S}_0$ ground state and the $^3\text{P}_0$ clock state~\cite{Hoefer2015,Pagano2015}. 
Since the defect is centered at the origin, the effective potential preserves spatial parity and couples only even-parity Fock states, thus preserving the $\mathbb{Z}_2$ symmetry required by our theoretical model. 

The dissipation processes of the QvdPo can be simultaneously engineered via continuous two-photon Raman sideband transitions~\cite{PoyatosCiracZoller1996,Li2025}, addressing two long-lived nuclear spin states of the primary atom's ground states: $|g\rangle$ and $|e\rangle$. Single-phonon gain with rate $\gamma_1$ is driven by the first-order blue motional sideband ($|g, n\rangle \to |e, n+1\rangle$). Instead, non-linear two-phonon losses with rate $\gamma_2$ are driven by the second-order red sideband ($|g, n\rangle \to |e, n-2\rangle$).  
The subsequent optical repumping cycle returns the atom to $|g\rangle$, making the engineered transitions effectively irreversible and realizing the desired time-independent Lindblad dissipators $\mathcal{D}[\hat{a}^\dagger]$ and $\mathcal{D}[\hat{a}^2]$. 

The simultaneous, continuous application of the Raman drives and optical pumping, together with the presence of the impurity-atom, is expected to cause  the primary atom's motional state to autonomously relax to the predicted phase-locked, Wigner-negative steady state $\hat{\rho}_\infty$. 

Some considerations are relevant to the experimental implementation of the proposed scheme. 
First, in order to operate in the Lamb-Dicke regime despite the large zero-point spatial spread associated with the shallow trap, the two Raman beams can be aligned with a small relative crossing angle. This decouples the momentum kick from the optical wavelength, thereby reducing the momentum transfer and the associated Lamb-Dicke parameter compared to the conventional counter-propagating Raman beam configuration. 

Second, the localized nature of the defect requires sufficient control over the impurity-atom's position. In particular, fluctuations in the tweezer's position should remain small compared to the zero-point spatial spread, so that the effective potential remains approximately parity symmetric. 

Finally, photon recoil during optical repumping can lead to motional heating, potentially destroying the steady-state Wigner negativity. This effect can be mitigated by directing the repumping beam along a transverse, tightly confined direction $\omega_\perp \gg \omega$ or by using sufficiently narrow optical transitions. The experimental requirement is that the recoil heating rate remains small with respect to the dissipative rates $\gamma_1$ and $\gamma_2$.

\section{Conclusions}

In this article, we have shown that inserting a parity-preserving localized defect provides an autonomous mechanism to stabilize non-classical steady states in a QvdPo, a paradigmatic driven-dissipative quantum system. 

By breaking the continuous $U(1)$-symmetry down to a discrete $\mathbb{Z}_2$ parity-symmetry, the defect protects even-parity coherences, with the lowest-order coherence $\rho_{02}$ playing a central role in the deep quantum regime. In fact, in such a regime, this coherence simultaneously gives rise to negative regions in the system's WF at the steady-state and enforces second-harmonic phase-locking, which freezes the QvdPo into two macroscopic, opposite phases. Moreover, an analytical threshold for the onset of Wigner negativity, derived within the Fock $n\leq 2$-truncated manifold, captures the numerical results across a broad range of system parameters. 

Although QvdPo dynamics have been experimentally realized in trapped-ion systems~\cite{Li2025}, implementing the localized defect considered here requires a different experimental strategy. Therefore, we propose an alternative platform based on ultracold alkaline-earth-like atoms confined in optical tweezers. In this setting, a pinned impurity atom provides an effective contact potential~\cite{Busch1998}, while Raman sideband transitions and optical repumping are used to engineer the single-phonon gain and two-phonon losses. This platform offers a promising route towards investigating defect-induced non-classicality and phase-locking in a driven-dissipative quantum system. 

Our results open several directions for future research. Extending the proposed mechanism to quantum spin van der Pol oscillators~\cite{KatoNakao_QSvdP_2025} could provide a natural setting for exploring defect-induced symmetry breaking in finite-dimensional collective spin systems. More broadly, the stabilization of non-classical states may open new opportunities for quantum information processing~\cite{Cacciapuoti2020} and quantum metrology~\cite{Giovannetti2011,Pezze2018}, as well as quantum memories~\cite{Hammerer2010,Zhu2025} and quantum energy storage systems~\cite{CampaioliRMP2024}. Moreover, arrays of coupled oscillators with localized defects could enable the study of higher-order quantum synchronization and driven-dissipative phase transitions. 

\begin{appendix}
\numberwithin{equation}{section}

\section{Unperturbed exact solution of the QvdPo}
\label{app:unperturbed_exact_solution}

In the absence of the defect, the Liouvillian $\mathcal{L}_0$ has a continuous $U(1)$ rotational symmetry. As a consequence, the steady state is diagonal in the Fock basis: $\hat{\rho}_\infty^{\rm (k=0)} = \sum_{n=0}^\infty p_n |n\rangle\!\langle n|$. The master equation for the populations $p_n$, defined in \eqref{eq:master_eq} of the main text, reduces to the recurrence relation
\begin{eqnarray}
    \label{eq:app_recurrence}
    \dot{p}_n = \gamma_1 n p_{n-1} + \gamma_2 (n+1)(n+2)p_{n+2} \nonumber \\
    - [\gamma_1(n+1) + \gamma_2 n(n-1)] p_n = 0,
\end{eqnarray}
where $\gamma_1$ and $\gamma_2$ are the single-photon gain and two-photon dissipation rates, respectively. To determine the populations $p_n$, we introduce the generating function $F(z) = \sum_{n=0}^\infty p_n z^n$. Multiplying \eqref{eq:app_recurrence} by $z^n$ and summing over $n$, yields the following second-order linear differential equation:
\begin{equation}\label{eq:diff_eq_F_z}
    (1-z^2) F''(z) + r z (1-z) F'(z) + r (1-z) F(z) = 0,
\end{equation}
where $r := \gamma_1/\gamma_2$ and the prime at the superscript denotes the derivative with respect to $z$. For $z \neq 1$, this simplifies to:
\begin{equation}
    (1+z) F''(z) - r z F'(z) - r F(z) = 0.
\end{equation}
Applying the variable transformation $x = r(1+z)$ and defining $G(x) \equiv F(\frac{x-r}{r})$, \eqref{eq:diff_eq_F_z} takes the form of a Kummer's confluent hypergeometric equation: 
\begin{equation}\label{eq:diff_eq_G_x}
    x G''(x) + (r-x) G'(x) - G(x) = 0.
\end{equation}
The general solution is given by $G(x) = C_1 \Phi(1, r, x) + C_2 U(1, r, x)$, where $\Phi$ and $U$ are the confluent hypergeometric functions of the first and second kind, respectively. Requiring the solution of \eqref{eq:diff_eq_G_x} to be regular at the origin $x=0$ and imposing the normalization condition $F(1) = \sum_n p_n = 1$, the solution of \eqref{eq:diff_eq_F_z} for $z \neq 1$ is
\begin{equation}
    F(z) = \frac{\Phi(1, r, r(1+z))}{\Phi(1, r, 2r)}.
\end{equation}

The individual populations $p_n$ are extracted by evaluating the $n$-th derivative of the generating function $F(z)$ at $z=0$:
\begin{equation}
    p_n = \frac{1}{n!} \frac{d^n F(z)}{dz^n} \bigg|_{z=0}.
\end{equation}
Using the identity $\frac{d^n}{dx^n} \Phi(a, b, x) = \frac{(a)_n}{(b)_n} \Phi(a+n, b+n, x)$, where $(a)_n := \Gamma(a+n)/\Gamma(a)$ is the Pochhammer symbol, we arrive at the exact analytical expression for $p_n$~\cite{Lorch2016, Dodonov1997}:
\begin{equation}
    p_n = \frac{r^n}{(r)_n} \frac{\Phi(1+n, r+n, r)}{\Phi(1, r, 2r)}\,,
    \label{eq:app_unpert_pop}
\end{equation}
which is \eqref{eq:pn_unpert} of the main text. These populations describe the occupation of the Fock states in the limit cycle developed by the QvdPo in the phase space. In fact, in the semiclassical limit ($\gamma_1 \gg \gamma_2$), the populations peak at $n \approx r$, which gives rise to the characteristic ring structure. 

\section{Matrix elements of the delta-shaped potential}
\label{app:Vmn}

To represent the delta-shaped potential $\hat{V} = k\delta(\hat{x})$ in the Fock basis, we project the operator $\hat{V}$ onto the eigenstates $\{|\phi_n\rangle\}$ of the unperturbed quantum harmonic oscillator. In this basis, the Hamiltonian separates into the unperturbed diagonal terms and a dense perturbation matrix resulting from the Dirac-delta function centered at the origin:
\begin{equation} 
    \left(\hat{H}_k \right)_{mn} = \omega\left(n + \frac{1}{2}\right)\delta_{mn} + V_{mn}, 
\end{equation}
where the matrix elements of the potential are given by $V_{mn} := k\phi_m(0)\phi_n(0)$. Due to the parity symmetry of the oscillator, all odd-parity eigenstates have a vanishing wave-function at the center of the trap, meaning $\phi_{2n+1}(0) = 0$ for any integer $n$. On the other side, the even-parity eigenstates at $x=0$ can be different than zero and evaluated analytically:
\begin{equation}
    \phi_{2n}(0) = (-1)^n \left(\frac{\omega}{\pi}\right)^{1/4} \frac{\sqrt{(2n)!}}{2^n n!} \,.
\end{equation}
Because $V_{mn} \neq 0$ only when both $m$ and $n$ are even, the delta-shaped potential only couples even-parity states. Furthermore, since the wave-functions at the origin are real-valued, the matrix is strictly symmetric and real: $V_{mn} = V_{nm} \in \mathbb{R}$. 

\section{Steady-state dynamics in the truncated manifold $n\leq 2$ and analytical threshold $k_{\rm crit}$}
\label{app:truncation_calculations}

To derive the steady-state in the deep quantum limit, we restrict the Fock space to the manifold $n \leq 2$. Using the master equation defined in \eqref{eq:master_eq}, the time-evolution of the populations $p_n$ and the leading off-diagonal coherence $\rho_{02}$ in the presence of $V_{mn} = \langle m | k \delta(\hat{x}) | n \rangle$ (see Appendix \ref{app:Vmn}) is given by the coupled system of equations
\begin{align}
    \dot{p}_0 &= -\gamma_1 p_0 + 2\gamma_2 p_2 + 2 V_{02} \Im\{\rho_{20}\}\label{eq:dot_p_0} \\
    \dot{p}_1 &= \gamma_1 p_0 - 2\gamma_1 p_1 \\
    \dot{\rho}_{02} &= -(\Gamma_{02} + i\Delta_{02})\rho_{02} - i V_{02}(p_2 - p_0),\label{eq:dot_rho_02}
\end{align}
where we have enforced the truncation by setting $p_{3} = 0$. The dissipative rates derived from the Lindblad superoperators yield the effective decoherence rate $\Gamma_{02} = 2\gamma_1 + \gamma_2$. Instead, $\Delta_{02}$ is the energy detuning that comprises the diagonal shifts induced by the potential:
\begin{equation}
    \Delta_{02} = -2\omega + V_{00} - V_{22} = \frac{k}{2}\sqrt{\frac{\omega}{\pi}} - 2\omega \,.
\end{equation}
Setting the time derivatives to zero in (\ref{eq:dot_p_0})-(\ref{eq:dot_rho_02}), the equation for $p_1$ immediately yields $p_1 = p_0/2$. Then, the probability conservation $p_0 + p_1 + p_2 = 1$ determines the second-excited state population as $p_2 = 1 - \frac{3}{2}p_0$.

The stationary condition for the coherence, $\dot{\rho}_{02}=0$, in (\ref{eq:dot_rho_02}) is
\begin{equation}\label{eq:rho02}
    \rho_{02} = \frac{-i V_{02}}{\Gamma_{02} + i\Delta_{02}} (p_2 - p_0),
\end{equation}
which corresponds to an adiabatic elimination mechanism if a number of levels larger than two were considered. \eqref{eq:rho02} implies that $\Im\{\rho_{20}\} = \frac{\Gamma_{02} V_{02}}{\Gamma_{02}^2+\Delta_{02}^2} (p_2 - p_0)$. Substituting this back into (\ref{eq:dot_p_0}) and introducing the coherent transfer rate $C_{02} := \frac{\Gamma_{02} V_{02}^2}{\Gamma_{02}^2 + \Delta_{02}^2}$ yields:
$0 = -\gamma_1 p_0 + 2\gamma_2 (1 - 3p_0/2) + 2 C_{02} (1 - 5p_0/2)$. Solving for $p_0$, we obtain the exact stationary populations as a function of $\gamma_1, \gamma_2, C_{02}$:
\begin{equation}\label{eq:p_steady_state}
    p_0 = \frac{2\gamma_2 + 2C_{02}}{\gamma_1 + 3\gamma_2 + 5C_{02}}, \quad p_1 = \frac{p_0}{2}, \quad p_2 = 1 - \frac{3}{2}p_{0}.
\end{equation}
In the limit of unperturbed system ($k \to 0$ that implies $C_{02} \to 0$), \eqref{eq:p_steady_state} precisely returns the steady-state value $p_0 \sim 2\gamma_2 / (\gamma_1 + 3\gamma_2)$.

To determine the critical strength $k_{\mathrm{crit}}$ of the potential, we use the condition for the onset of Wigner negativity, $\sqrt{2} |\rho_{02}| > p_1 - 2p_2 + \sqrt{2p_2(p_0 - p_1 + p_2)}$ [\eqref{eq:NegativityCondition} of the main text], and we substitute the coherence amplitude 
\begin{equation}\label{eq:modulus_rho_02}
    |\rho_{02}| = \frac{\left|p_2 - p_0\right|\,\left|V_{02}\right|}{\sqrt{\Gamma_{02}^2 + \Delta_{02}^2}}
\end{equation}
obtained from the truncated steady-state solution [\eqref{eq:rho02}]. Solving this condition gives the threshold value of $k$ above which the WF becomes negative. For this purpose, we consider the parameter regime where $\gamma_2 \gg \gamma_1, k, \omega$. This entails that $\Gamma_{02} \gg \Delta_{02}$ such that $C_{02}$ is vanishing and $p_0 \approx 2/3$, $p_1 \approx 1/3$ and $p_2 \approx r/3$ with $r = \gamma_1/\gamma_2$ infinitesimal. As a result,
\begin{equation}\label{eq:approx_pops}
    p_1 - 2p_2 + \sqrt{2p_2(p_0 - p_1 + p_2)} \approx \frac{1 - 2r + \sqrt{2r(1+r)}}{3}.
\end{equation}
Then, substituting in (\ref{eq:modulus_rho_02}) the matrix element $|V_{02}| = \frac{k}{\sqrt{2}} \sqrt{\frac{\omega}{\pi}}$, the decoherence rate $\Gamma_{02} = \gamma_2(1+2r)$, and the approximation $\Delta_{02} \approx -2\omega$ of the detuning in the weak-coupling, we find:
\begin{equation}\label{eq:approx_rho_02}
    \sqrt{2}|\rho_{02}| \approx k \sqrt{\frac{\omega}{\pi}} \frac{2-r}{3 \sqrt{4\omega^2 + \gamma_2^2(1+2r)^2}}.
\end{equation}
Finally, equating the right-hand side of Eqs.~\ref{eq:approx_pops}-\ref{eq:approx_rho_02} and isolating $k$ yields the boundary of \eqref{eq:k_crit_analytical} in the main text:
\begin{equation}
    k_{\rm crit} = \sqrt{\frac{\pi}{\omega}} \sqrt{4\omega^2 + \gamma_2^2(1+2r)^2} \left[ \frac{1 - 2r + \sqrt{2r(1+r)}}{2-r} \right].
\end{equation}

\section{Quantitative justification of the ($n\leq2$)-manifold truncation}
\label{app:Truncation}

\begin{figure}
    \centering
    \includegraphics[width=0.85\linewidth]{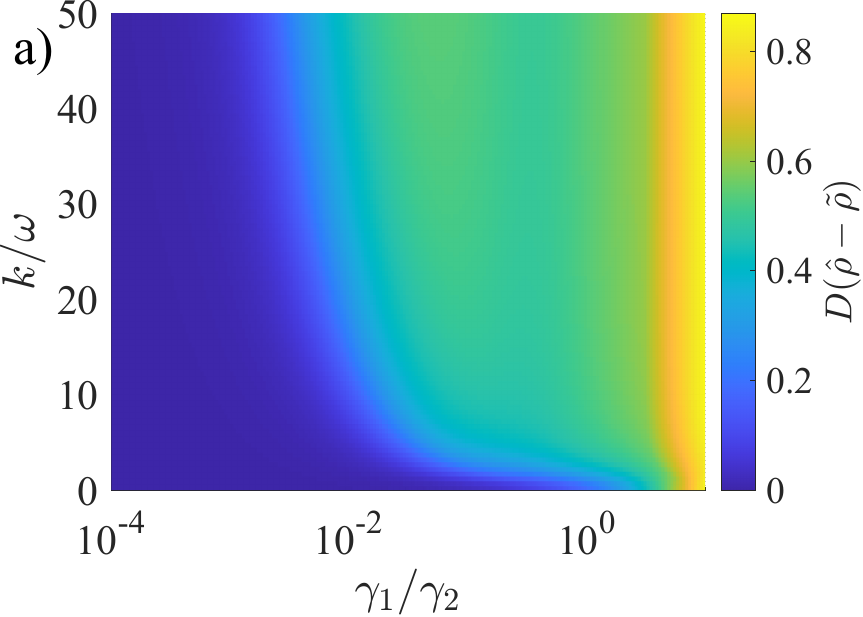} \\
    \includegraphics[width=0.85\linewidth]{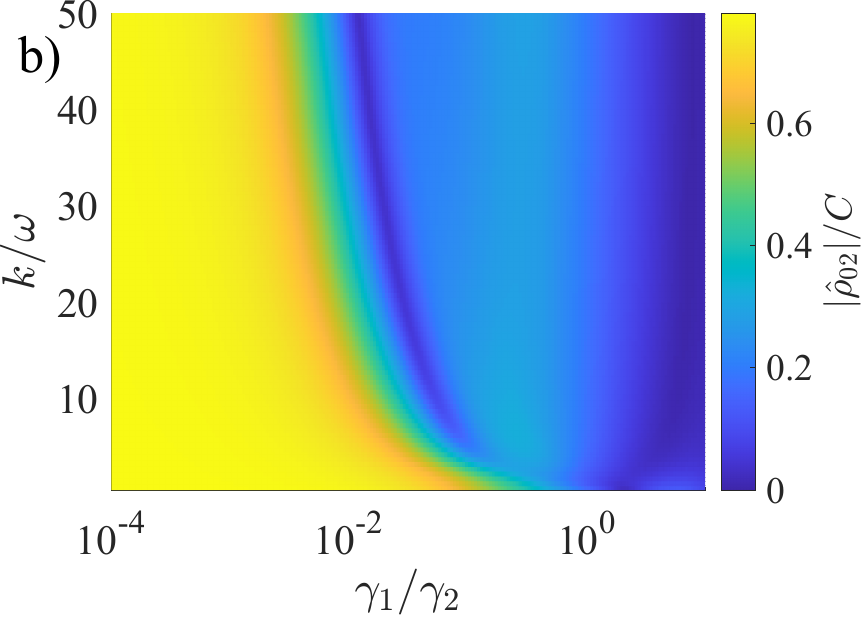}
    \caption{
    (a) Trace distance $D(\hat{\rho}, \widetilde{\hat{\rho}})$ between the complete numerical density matrix $\hat{\rho}$ and the density matrix $\widetilde{\hat{\rho}}$ derived in the truncated $(n\leq 2)$-manifold. (b) Ratio between the absolute value of the lowest-order coherence $\rho_{02}$ and the $\ell_1$-norm of coherence $\mathcal{C}$.
    }
    \label{fig:TraceDistance}
\end{figure}
 
To assess the validity of the truncation to the ($n\leq 2$)-manifold, we compare the truncated steady state $\widetilde{\hat{\rho}}$ with the complete numerical steady state $\hat{\rho}$, obtained using the exact Green-superoperator formalism. Fig.~\ref{fig:TraceDistance}(a) shows the trace distance between these two states:
\begin{equation}
    D\left( \hat{\rho}, \widetilde{\hat{\rho}} \right) = \frac{1}{2} \mathrm{Tr} \sqrt{ \left( \hat{\rho} - \widetilde{\hat{\rho}}\right)^{\dagger}\left( \hat{\rho} - \widetilde{\hat{\rho}} \right)},
\end{equation}
In the deep quantum regime ($\gamma_1/\gamma_2 \ll 1$), $D(\hat{\rho}, \widetilde{\hat{\rho}})$ remains small for weak potential strength $k/\omega$. For $k > k_{\text{crit}}$, the discrepancy increases and the WF is split into two lobes; then, describing the full WF profile requires contributions from higher even-parity states ($n = 4, 6, \dots$). 
Nevertheless, the interference pattern responsible for Wigner negativity is primarily governed by the lowest-order parity-preserving coherence, $\rho_{02}$. To quantify its contribution, in Fig.~\ref{fig:TraceDistance}(b) we show the ratio between $|\rho_{02}|$ and the $\ell_1$-norm of coherence: $|\rho_{02}|/\mathcal{C}$, where $\mathcal{C} := \sum_{i \neq j} |\rho_{ij}|$. In the deep quantum regime, this ratio remains at $\approx 0.6 - 0.7$, indicating that $\rho_{02}$ is the dominant coherence at the steady-state regardless of $k$. As $\gamma_1/\gamma_2$ approaches or exceeds unity, higher-order coherences, including $\rho_{24}$ and $\rho_{46}$, become increasingly important. The resulting decrease in the relative weight of $|\rho_{02}|$ signals the breakdown of the truncated manifold approximation and the need for additional Fock levels. 

\section{Quantum phase operator \& mean resultant length}
\label{app:MRL}

We introduce the Susskind-Glogower phase operator in the infinite-dimensional limit~\cite{Susskind1964,PeggBarnett1989}:
\begin{equation}
   \hat\chi = \sum_{n=0}^\infty |n\rangle\!\langle n+1|.
\end{equation}
The first-harmonic Mean Resultant Length (MRL) is defined as $R_1 = |\langle \hat\chi \rangle|$. For any state $\hat\rho$ respecting the parity symmetry ($[\hat{\rho}, \hat{\Pi}] = 0$), the first-harmonic phase operator $\hat\chi$ is odd under parity transformations ($\hat{\Pi} \hat{\chi} \hat{\Pi}^\dagger = -\hat{\chi}$). Therefore, its expectation value vanishes, so that $R_1=0$.  
The lowest non-vanishing moment of $\hat{\chi}$ is the second-harmonic, whose MRL is
\begin{equation}
    R_2 = \left| \left\langle \hat{\chi}^2 \right\rangle \right| = \left| \mathrm{Tr} \left[ \hat{\rho} \sum_{n=0}^\infty |n\rangle\!\langle n+2| \right]\right|.
\end{equation}
Within the truncated manifold with $n \le 2$, the second-harmonic phase operator reduces to $\hat{\chi}^2 = |0\rangle\!\langle 2|$, yielding $R_2 = |\rho_{02}|$ at the steady-state. 
Hence, $R_2$ provides a measure of second-harmonic phase localization, and even synchronization, that is compatible with parity symmetry.

\end{appendix}

\bibliography{bibliography}

\end{document}